\documentclass[a4paper]{mem}
\usepackage{natbib}
\usepackage{graphicx}
\usepackage[a4paper]{hyperref}
\idline{73}{23}
\begin{document}
\title{ROSAT-SDSS Galaxy Clusters Survey.} 
\subtitle{Correlating X-ray and optical properties.}
\author{P. Popesso, H. B\"ohringer, W. Voges}
\institute{ Max-Planck-Institut fur extraterrestrische Physik, 85748 Garching, Germany}
\authorrunning{P. Popesso et al.}
\abstract{
For a detailed comparison of the appearance  of cluster of galaxies in
X-rays and  in the optical,  we have compiled a comprehensive database
of X-ray and optical  properties of a sample  of clusters based on the
largest available X-ray and optical surveys:  the ROSAT All Sky Survey
(RASS) and the Sloan Digital Sky  Survey (SDSS).  The RASS-SDSS sample
comprises 114  X-ray selected   clusters.  For  each system   we  have
uniformly  determined the X-ray  and    optical properties .  For    a
subsample of 53 clusters we have also compiled the temperature and the
iron abundance from the literature.  The  total optical luminosity can
be  determined   with a typical  uncertainty  of  20\%  with  a result
independent of the choice  of local or global  background subtraction.
We searched for parameters which provide  the best correlation between
the X-ray  luminosity and the optical  properties and found that the z
band  luminosity  determined within  a  cluster  aperture  of  0.5 Mpc
$\rm{h}_{70}^{-1}$ provides   the best correlation  with  a scatter of
about 60\%.  The resulting correlation of $L_X$ and  $L_{op}$ in the z
and i bands shows a logarithmic slope  of 0.38, a value not consistent
with  the assumption  of a   constant $M/L$.    Consistency is  found,
however, for an increasing  M/L with luminosity.  This is confirmed by
our analysis of the cluster mass to light ratio for  a subsample of 53
clusters. We found a significant  dependence  of $M/L$ on the  cluster
mass with a logarithmic  slope ranging from 0.27  in the i and r bands
to 0.22 in the z band.
 }
\authorrunning{P. Popesso et al.}
\titlerunning{ROSAT-SDSS Galaxy Clusters Survey.}
\maketitle

\section{Introduction}
Cluster  of galaxies are the largest  well  defined building blocks of
our
 Universe. They form via  gravitational collapse of cosmic
 matter
over a region of several megaparsecs. Cosmic baryons, which
 represent
approximately  10-15\%  of the  mass content of  the Universe,
 follow
dynamically the dominant dark matter during the collapse. As a
 result
of  adiabatic  compression and   of  shocks generated  by  supersonic
motions,  a    thin hot  gas   permeates   the  cluster gravitational
potential.  For a typical cluster  mass  of $10^{14}$ $M_{\odot}$ the
intracluster gas reaches a temperature of the order of $10^7$ keV and,
thus, radiates   optically  thin    thermal bremsstrahlung and    line
radiation in  the X-ray band.
In addition  to  the   hot,  diffuse  component,   baryons   are also
concentrated in the individual galaxies within the cluster. These are
best studied through photometric  and spectroscopic  optical surveys,
which provide  essential   information about luminosity,   morphology,
stellar
 population and  age. 

Since  solid observational  evidences  indicate  a strong
 interaction
between the   two  baryonic  components, one  of    our goals is  the
comparison  of the X-ray and  the  optical appearance of the clusters.
We want in particular find optical parameters that provide the closest
correlation to  the X-ray parameters, such that  we can predict within
narrow  uncertainty  limits the  X-ray  luminosity  from these optical
parameters  and viceversa.  On the basis   of these considerations, we
have created a  large database
 of  clusters of galaxies  based on the
largest available X-ray and
 optical surveys: the ROSAT All Sky Survey
(RASS),  and the  Sloan   Digital Sky  Survey  (SDSS).    By carefully
combining the data of  the two surveys  we
 have created the RASS-SDSS
galaxy cluster  catalog.  The X-ray-selected
  galaxy clusters cover a
wide  range  of  masses, from groups   of
  $10^{12.5}$ $M_{\odot}$ to
massive clusters of $10^{15}$   $M_{\odot}$
 in a redshift  range from
0.002 to 0.45.  The RASS-SDSS sample comprises  all the X-ray selected
objects   already  observed in  the sky   region covered  by the Sloan
Digital  Sky
  Survey.   For  all  derived   quantities, we  have used
$\rm{H}_0=70$ $\rm{km}$$\rm{s}^{-1}$$\rm{Mpc}^{-1}$, $\Omega_ {m}=0.3$
and $\Omega_ {\lambda}=0.7$.

\section{The data.}
In order to create a homogeneous  catalog of X-ray cluster properties,
we  have calculated all X-ray parameters  using only RASS data for all
clusters in the sample.  The X-ray luminosity has been calculated with
the growth  curve analysis (GCA) method used  for REFLEX and  NORAS 2,
based on  the RASS3 database  (B\"ohringer et al., 2001).  The optical
photometric  data  were taken from   the SDSS (York   et  al. 2000 and
Stoughton  et al.  2002).  The SDSS  consists  of an imaging survey of
$\pi$ steradians of the northern sky in the five passbands u, g, r ,i,
z, in the entire optical range.  Since the galaxies  do not have sharp
edges  or a  unique surface  brightness profile,  it  is nontrivial to
define   a  flux  for  each  object.  The    SDSS photometric pipeline
calculates  a number of  different  magnitudes for each  object: model
magnitude,   Petrosian magnitudes  and  PSF  magnitudes.   In the data
analysis of this paper  we used the  Petrosian magnitudes for galaxies
brighter then 20  mag and the psf magnitudes  for objects fainter than
20 mag.

\section{Optical Luminosity from SDSS data}
%-------------------------------------------------------------
   \begin{figure*}
   \centering
   \resizebox{\hsize}{!}{\includegraphics[clip=true]{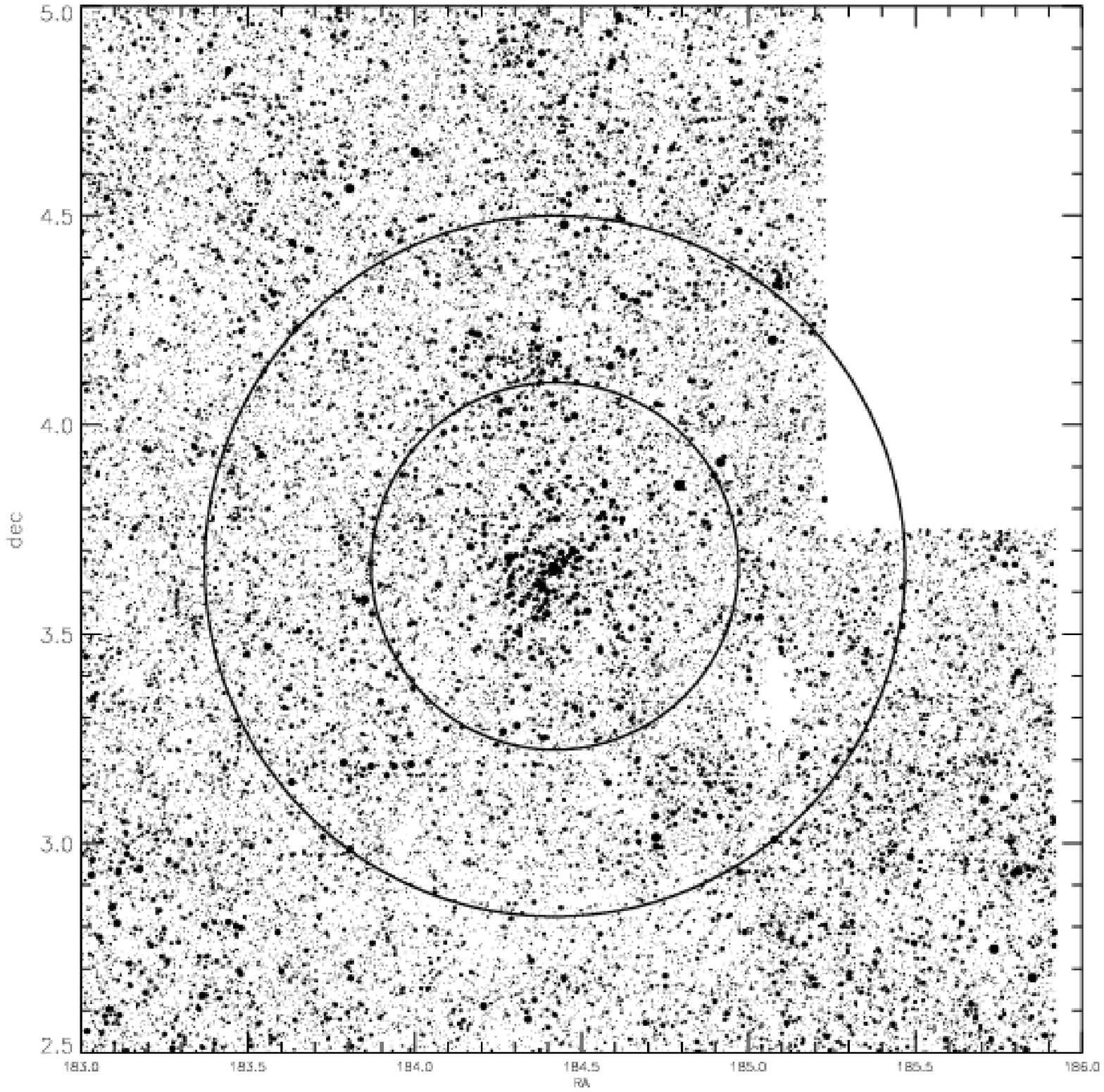}
   \includegraphics[clip=true]{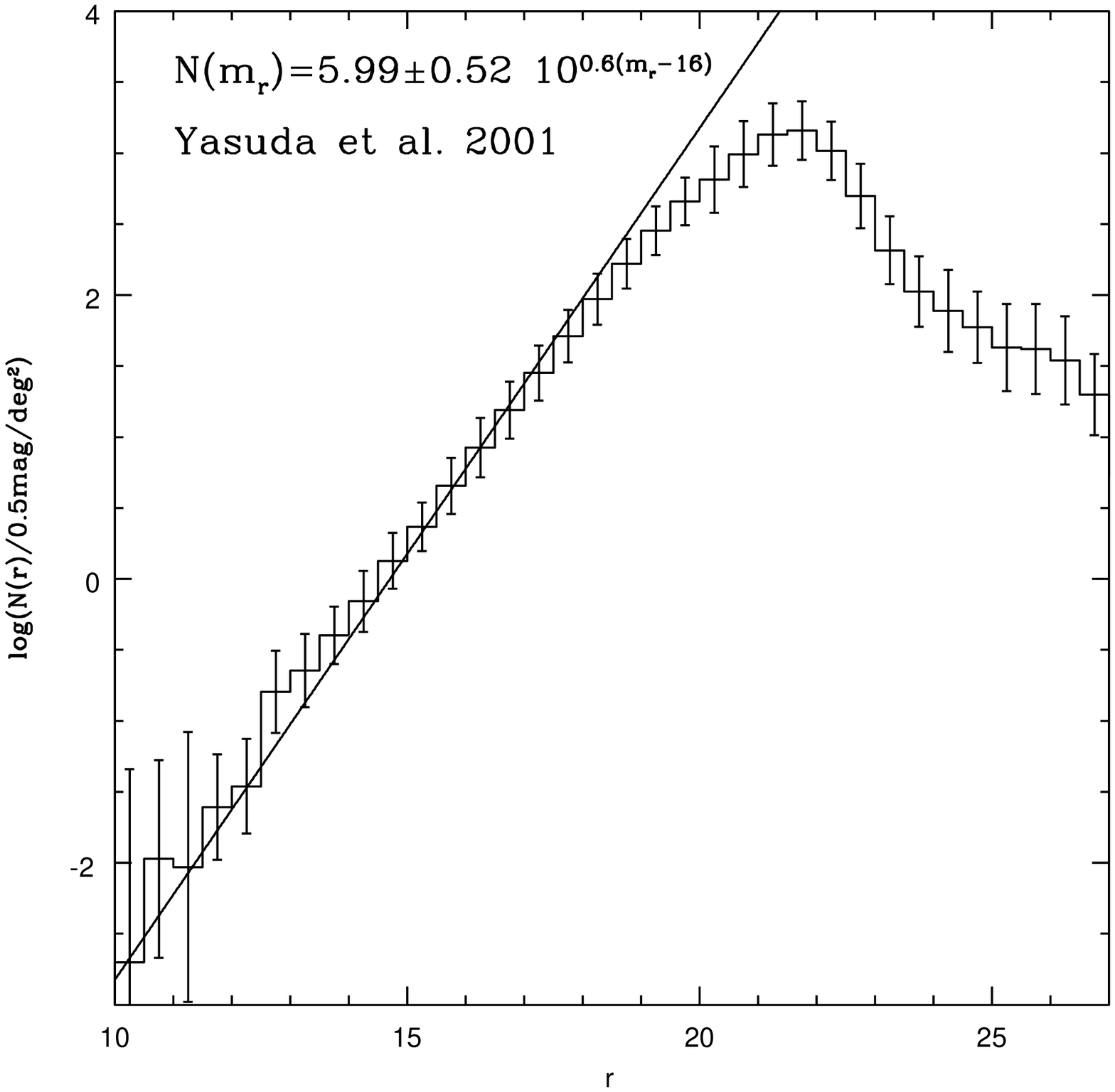}}
\caption{
Left panel: The plot shows a cluster region with the local background.
The  dots represent the   galaxies in  the sample.   The  biggest dots
correspond  to the  brightest galaxies in   apparent r magnitude.  The
local background number counts have been calculated inside the annulus
with inner radius   equal to $r_{200}+0.2$ deg   and a width  of $0.5$
degree. The regions with  voids due to  lack of data or close clusters
have to be discarded in the background estimation. Right panel: Global
background   number counts  as  a  function of  magnitude  in  the $r$
band.   The  error bars include    the  contribution  of  large  scale
structure. The line shows the counts-magnitude relation expected for a
homogeneous galaxy distribution in a universe with euclidian geometry:
$N(r)=A_r10^{0.6(r-16)}$.            The          value             of
$A_r=5.99\pm0.52(0.5mag)^{-1}deg^{-2}$  is  the results of the  fit in
Yasuda et al. 2001 for $12\le r \le 17$.}
\label{local}
\end{figure*}
%
%_____________________________________________________________

The total optical luminosity  of a cluster has  to be calculated after
the subtraction     of    the  foreground   and     background  galaxy
contamination. Since we have used  only photometric data from the SDSS
galaxy catalog, we  have no direct information  on  the cluster galaxy
memberships.   There are  two  different approaches   to overcome this
problem:  a color  cut or  a statistical  background  subtraction.  We
chose the latter approach since the former method may introduce a bias
against bluer  cluster members observed  to have varying fractions due
to the Butcher-Oemler effect.

We have   considered  two  different  approaches to   the  statistical
subtraction of the  galaxy  background.  First  we have   calculated a
local background within an    annulus centered on the cluster    X-ray
center. The   annulus has been then divided   in 20 sectors  and those
featuring   a larger than $3\sigma$ deviation   from the median galaxy
density are discarded from  the further calculation (fig. \ref{local},
left panel).

As a second method we have derived a global background correction. The
galaxy number counts $N_{bg}^g(m)dm$ was  derived from the mean of the
magnitude number counts determined in five different SDSS sky regions,
each with an  area of 30  $\rm{deg^2}$ (fig.\ref{local}, right panel).
The  source  of  uncertainty in  this  second case  is systematic  and
originates the presence   of large-scale clustering  within the galaxy
sample,  while the Poissonian error of  the galaxy counts is small due
to the large area involved.  After the background subtraction we found
that the signal to noise in the  u band was too low  to be useful, and
performed  our  analysis on  the 4  remaining Sloan  photometric bands
g,r,i,z.

In order to calculate the total cluster luminosity, we have calculated
first the absolute magnitude, corrected for  Galactic extinction
and  K-correction.  We have  estimated    $L$  by using  the
(background corrected) magnitude number counts of the cluster galaxies
with the following prescription:
\begin{equation}
L=\sum_{i}^{N}{N_i(m)l_i(m)}+\int_{m_{lim}}^{\infty}{\phi (m)dm}
\end{equation}
The sum on right  side is performed  over  all the $N$  magnitude bins
with galaxy number $N_i(m)$ and mean luminosity $l_i(m)$. The integral
is an incompleteness correction  due to the  completeness limit of the
galaxy sample at $m_{lim}=21$ mag in the five Sloan photometric bands.
$\phi (m)$ is  the individual Schechter  luminosity function fitted to
the galaxy sample of each cluster. The incompleteness correction is of
the order of 5-10\% in the whole cluster sample.   This means that the
galaxies  below  the   magnitude   limit do not    give  a significant
contribution to the   total  optical luminosity.  Therefore   the most
important source of error is due to the contribution of the background
galaxy  number counts.  The uncertainty  in  each bin of magnitude  is
given    by   the  Poissonian      error     of   the   bin     counts
($\sqrt{N^i_{tot}(m)}$,    with  $i=1,...,N$)   and   the   background
subtraction  in  each magnitude  bin ($\sigma_{bg}^{i}(m)$). Since the
galaxy counts in the bins are independent, the error in the luminosity
is given by:
\begin{equation}
\Delta L=(\sum_{i}^{N}{(N^i_{tot}(m)+ \sigma_{bg}^{i}(m)^2)})^{\frac{1}{2}}.
\end{equation}

\section{Correlating X-ray and optical properties}

For  a cluster  in which  mass   traces optical light  ($M/L_{opt}$ is
constant), the gas is in hydrostatic equilibrium ($T \propto M^{2/3}$),
and $L_X \propto T^3$, we expect the X-ray bolometric luminosity to be
related to the optical luminosity as $L_{op}
\propto L_{X}^{0.5}$ and to the intracluster medium temperature as 
$L_{op} \propto T_{X}^{1.5}$. 

In this section  we show  that tight  correlations  exist between  the
total optical cluster luminosity and  the X-ray cluster properties  as
the X-ray luminosity and the intracluster medium temperature.

To  search  for  the   best  correlation  between  optical  and  X-ray
properties and  to optimally predict for example  the X-ray luminosity
from  the  optical  appearance,   we  are  interested  in  an  optical
characteristic,  which shows  a minimum  scatter in  the X-ray/optical
correlation. Therefore,  we performed  a correlation using  4 of  the 5
SDSS optical  band, $g$, $r$,  $i$ and $z$,  to find out  which filter
should  be  used in  the  prediction.  We  used  a  fixed aperture  to
calculate the optical luminosities for  all the clusters, to make no a
priori assumption  about the cluster size. Moreover,  to check whether
the scatter in the correlation depends on the cluster aperture, we did
the  same   analysis  with different apertures.   To quantify  the $L_{op} -
L_X$ and the $L_{op} - T_x $ relations, a linear regression in log-log
space was performed.
 In any photometric band the  scatter has a clear
dependence on the cluster aperture  by showing a  region of minimum on
the very center of the cluster, around 0.5 Mpc $\rm{h}_{70}^{-1}$.

%-------------------------------------------------------------
   \begin{figure*}
   \centering
   \resizebox{\hsize}{!}{\includegraphics[clip=true]{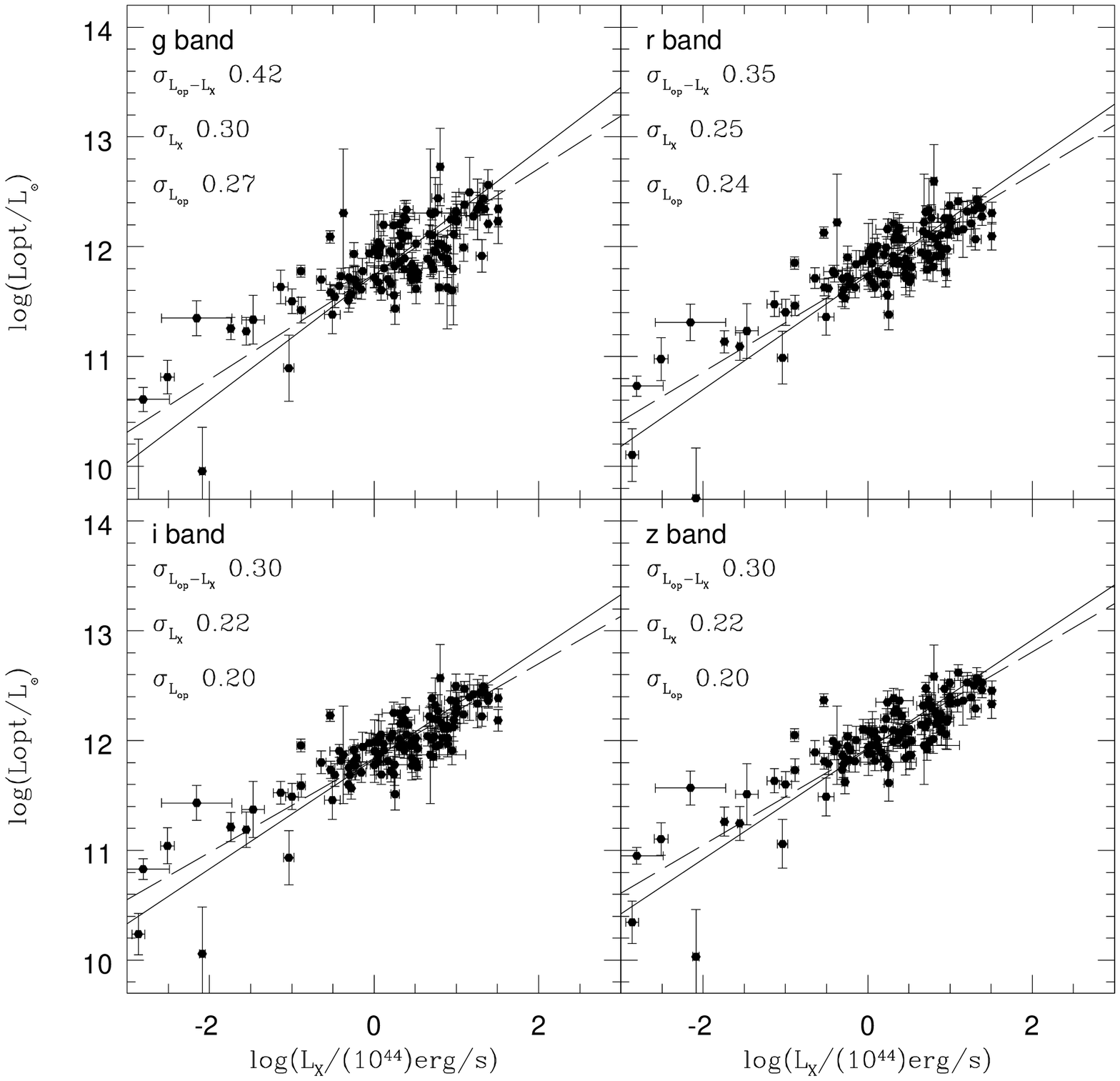}
   \includegraphics[clip=true]{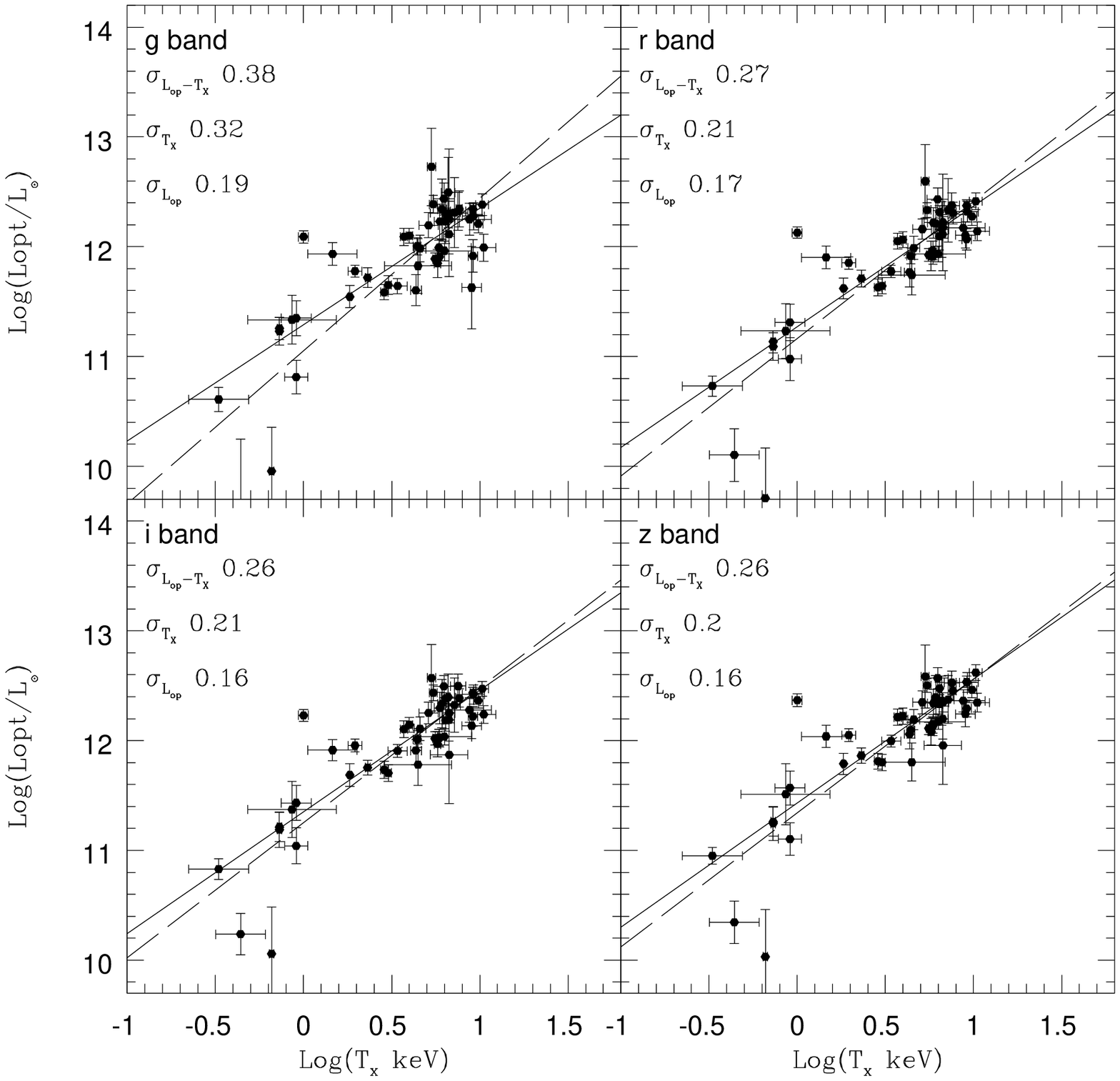}}
\caption{
Left   panel: Correlation  between    optical luminosities  and  X-ray
luminosities.  The  fit is performed  with a  linear regression in the
$log(L_{opt})-log(L_X)$ space   for each of the   4 optical bands. The
solid and   the dashed lines  are  the results  of the  orthogonal and
bisector   method  respectively.  The   error   bars are  at  the 68\%
confidence level in  both the variables.  Right  panel: the same as in
the left  panel  but for  the correlation  between  $L_X$ the  the ICM
temperature $T_X$}
\label{lx}
\end{figure*}
%
%_____________________________________________________________

Fig. \ref{lx} shows the $L_{op}  - L_X$ and $L_{op} - T_x
$ relation respectively, at the   radius of minimum scatter, 0.5  Mpc
$\rm{h}_{70}^{-1}$, in the z band. In fact the i  and the z bands have
a slightly  smaller scatter  than in the  other  optical bands at  any
radius. The best  fit in the z band  for the $L_{op} - L_X$  and the $L_{op} -
T_x $ relations at the radius of minimum scatter are respectively:
\begin{equation}
L_{op}/L_{\odot}=10^{11.79 \pm 0.02} L_X(_{Rosat})^{0.45\pm 0.03}
\end{equation}
\begin{equation}
L_{op}/L_{\odot}=10^{11.75 \pm 0.02} L_X(_{Bol})^{0.38\pm 0.02}
\end{equation}
\begin{equation}
L_{op}/L_{\odot}=10^{11.42 \pm 0.06} T_X^{1.12\pm 0.08}
\end{equation}

\section{Discussion and conclusions}
\begin{figure*}
   \centering
   \resizebox{\hsize}{!}{\includegraphics[clip=true]{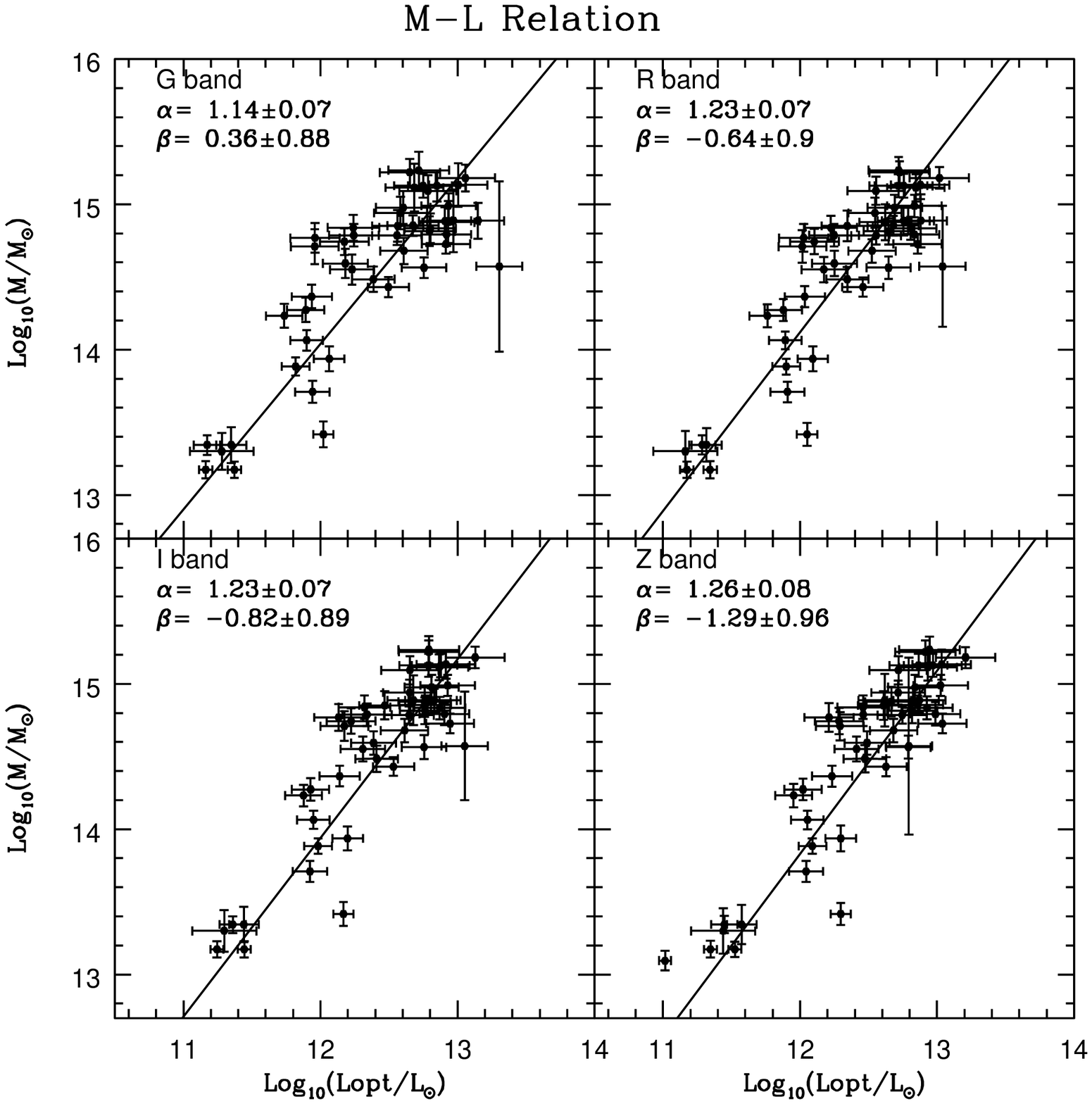}
   \includegraphics[clip=true]{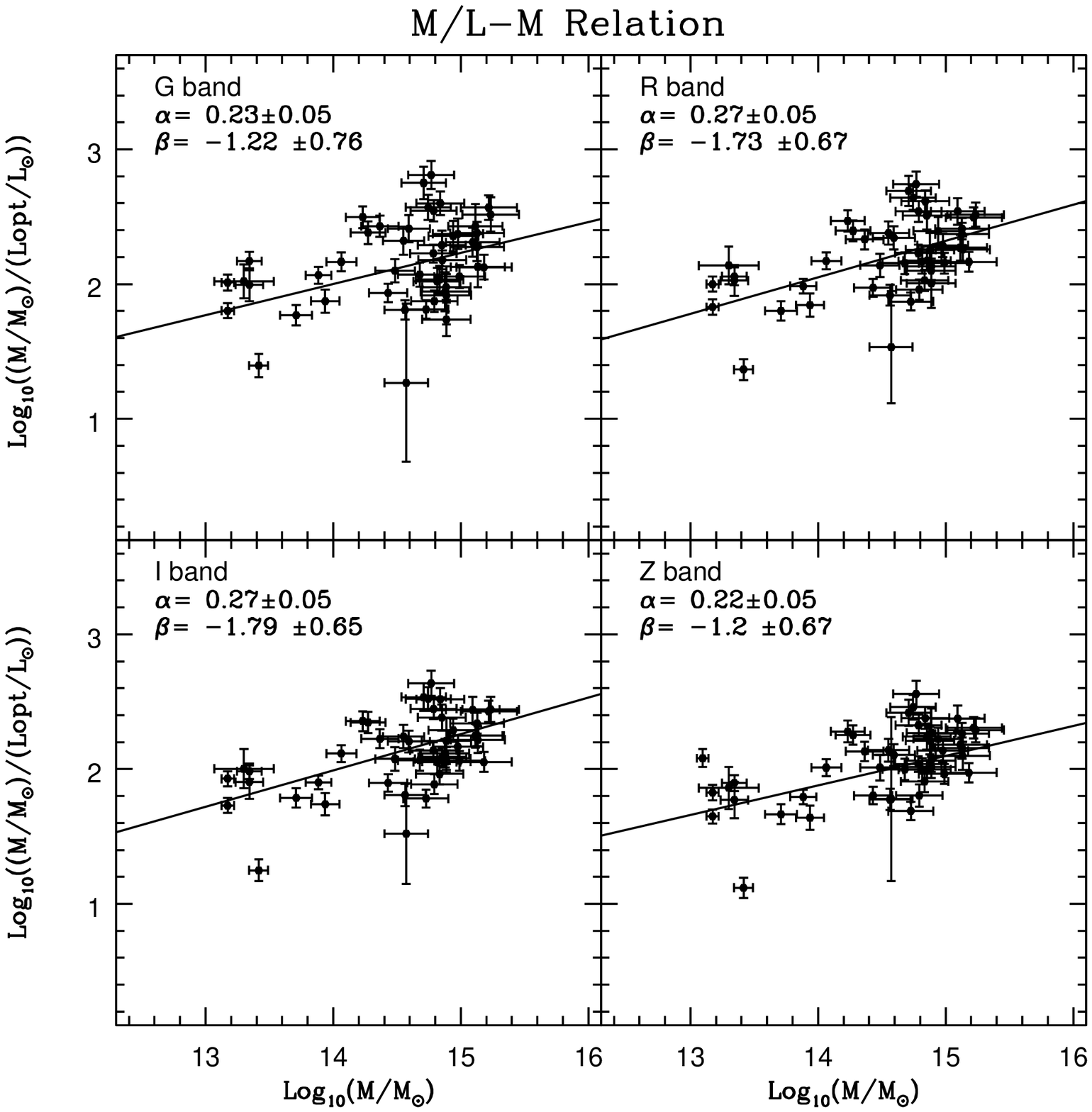}}
\caption{
Left panel: cluster mass within $r_{500}$ as a function of the optical
luminosity calculated  within  the same radius.  Right  panel: cluster
mass to light   ratio as a   function of the   cluster mass.   Obvious
consequence of  the $M-L_{op}$ relation  is  that $M/L$ shows  a clear
dependence on the cluster mass with a  slope from 0.27 in  the i and r
bands to 0.22 in the z band.  This is in agreement with the prediction
based on our new $L_X-L_{op}$ relation.  }
\label{ml}
\end{figure*}

The  value  of  the exponent   in the power   law  for the  $L_{opt} -
L_X(_{Bol})$ relation, is  around 0.38 in the  region of minimum.  The
values are not  consistent within  the errors  with  the value of  0.5
predicted under the assumption of hydrostatic equilibrium and constant
mass  to light  ratio.  The same  conclusion  can  be derived for  the
$L_{opt} - T_X$ relation. A simple reason of the disagreement could be
due  to the assumption of  a constant mass  to  light ratio.  In fact,
Girardi et  al.  (2002) analysed in detail  the mass to light ratio in
the B band of a sample of 294 clusters and groups , finding $M/L
\propto  L^{0.33\pm 0.03}$.  The  same  results was  found  by Lin  et
al.  (2003) in the  K band.  Thus, if we   consider this dependence of
$M/L$ from the optical luminosity  with the assumptions of hydrostatic
equilibrium, the new expected  relation between the optical luminosity
and the   X-ray    luminosity and  temperature  are   $L_{op}  \propto
L_X^{0.4}$ and $L_{op} \propto T_X^{1.25}$, respectively, which are in
good agreement  with  our results.  The  behavior  of the  $M/L$  as a
function of the  cluster mass and  luminosity is confirmed also by our
data as shown in fig. \ref{ml}. We used  the subsample of cluster with
known  temperature to calculate  the  mass within  $r_{500}$  with the
$M-T$  relation of  Finoguenov  et al.   (2001).  $M/L$ shows a  clear
dependence on the cluster mass with a  slope from 0.27 in  the i and r
bands to 0.22 in the z band.

The scatter in the $L_{op} - L_x $  relation for the aperture with the
best   correlation (0.5   Mpc  $\rm{h}_{70}^{-1}$),  in   the $L_{op}$
variable is 0.20, and the scatter in the $L_X$ variable is 0.22 in the
correlations obtained in i and the z bands.  Therefore, by calculating
the total cluster  luminosity in the  central part of the system,  one
can use the $i$ or  $z$ band to  predict the X-ray luminosity from the
optical data with a mean error of $60\%$.  In the same way the optical
luminosity      can   be  derived from         $L_X$   with the   same
uncertainty. Analogous results  are obtained for the  $L_{op}  - T_x $
relation. Since the observational uncertainties in  the optical and in
the X-ray  luminosity are  about 20\%, the  scatters of  60\% in  both
relations should be intrinsic.

To better  understand the relation  between the X-ray and  the optical
properties and their physical implications, the optical luminosity has
to be calculated within the physical size of the systems. This will be
discussed in a forthcoming publication.

%-------------------------------------------------------------
%
%_____________________________________________________________

Funding for the creation and distribution of the SDSS Archive has been
provided  by    the  Alfred P.   Sloan   Foundation, the Participating
Institutions, the  National Aeronautics and  Space Administration, the
National  Science  Foundation, the  U.S.    Department of  Energy, the
Japanese Monbukagakusho, and the Max Planck Society. The SDSS Web site
is   http://www.sdss.org/. The  SDSS is  managed  by the Astrophysical
Research Consortium   (ARC) for the   Participating Institutions.  The
Participating Institutions  are The  University of Chicago,  Fermilab,
the Institute for Advanced  Study, the Japan Participation  Group, The
Johns  Hopkins    University,  Los Alamos   National  Laboratory,  the
Max-Planck-Institute for   Astronomy (MPIA), the  Max-Planck-Institute
for   Astrophysics (MPA), New Mexico  State  University, University of
Pittsburgh, Princeton University, the United States Naval Observatory,
and the University of Washington.

\bibliographystyle{aa}

\begin{thebibliography}{}
\bibitem[{Alonso et al. (1996)}]{alonso}  Abazajian, K., Adelman, J., Agueros, M.,et al. 2003, AJ, 126, 2081 (Data Release One)
\bibitem[{Alonso et al. (1996)}]{alonso}  Blanton, M., Blanton, M. R., Dalcanton, J., Eisenstein, D., et al. 2001, AJ, 121, 2358;
\bibitem[{Alonso et al. (1996)}]{alonso}  Blanton, M.R., Lupton, R.H., Maley, F.M. et al. 2003, AJ, 125, 2276 (Tiling Algorithm)
\bibitem[{Alonso et al. (1996)}]{alonso}  B\"ohringer, H.,  Schuecker, P., Guzzo, L., et al. 2001, A\&A, 369, 826;
\bibitem[{Alonso et al. (1996)}]{alonso}  Eisenstein, D. J., Annis, J., Gunn, J. E., et al. 2001, AJ, 122, 2267;
\bibitem[{Alonso et al. (1996)}]{alonso}  Finoguenov, A., Reiprich, T. H., Böhringer, H. 2001,A\&A,368,749;
\bibitem[{Alonso et al. (1996)}]{alonso}  Fukugita, M., Ichikawa, T., Gunn, J. E. 1996, AJ, 111, 1748;
\bibitem[{Alonso et al. (1996)}]{alonso}  Girardi, M., Manzato, P., Mezzetti, M., et al. 2002, ApJ, 569, 720;
\bibitem[{Alonso et al. (1996)}]{alonso}  Gunn, J.E., Carr, M.A., Rockosi, C.M., et al 1998, AJ, 116, 3040 (SDSS Camera)
\bibitem[{Alonso et al. (1996)}]{alonso}  Hogg, D.W., Finkbeiner, D. P., Schlegel, D. J., Gunn, J. E. 2001, AJ, 122, 2129;
\bibitem[{Alonso et al. (1996)}]{alonso}  Lin, Y., Mohr, J. J., Stanford, S. A. 2003, ApJ,591,749;
\bibitem[{Alonso et al. (1996)}]{alonso}  Lupton, R. H., Gunn, J. E., Szalay, A. S. 1999, AJ, 118, 1406;
\bibitem[{Alonso et al. (1996)}]{alonso}  Lupton, R., Gunn, J. E., Ivezi\'c, Z.,  et al.  2001, (astro-ph/0101420);
\bibitem[{Alonso et al. (1996)}]{alonso}  Pier, J.R., Munn, J.A., Hindsley, et al. 2003, AJ, 125, 1559
\bibitem[{Alonso et al. (1996)}]{alonso}  Schlegel, D., Finkbeiner, D. P., Davis, M. 1998, ApJ, 500, 525;
\bibitem[{Alonso et al. (1996)}]{alonso}  Shimasaku, K., Fukugita, M., Doi, M., et al. 2001, AJ, 122, 1238;
\bibitem[{Alonso et al. (1996)}]{alonso}  Smith, J.A., Tucker, D.L., Kent, S.M., et al. 2002, AJ, 123, 2121;
\bibitem[{Alonso et al. (1996)}]{alonso}  Stoughton, C., Lupton, R.H., Bernardi, M., et al. 2002, AJ, 123, 485;
\bibitem[{Alonso et al. (1996)}]{alonso}  Strauss, M. A., M.A., Weinberg, D.H.,\bibitem[{Alonso et al. (1996)}]{alonso}  Lupton, R.H. et al. 2002, AJ, 124, 1810;
\bibitem[{Alonso et al. (1996)}]{alonso}  Yasuda, N., Fukugita, M. Narayanan, V. K. et al. 2001, AJ, 122, 1104;
\bibitem[{Alonso et al. (1996)}]{alonso}  York, D. G., Adelman, J., Anderson, J.E.,  et al. 2000, AJ, 120, 1579;

\end{thebibliography}

\end{document}